\documentclass[%
 reprint,                                                                                                                 
showpacs,preprintnumbers,                                                                                                 
 amsmath,amssymb,                                                                                                         
 aps,                                                                                                                     
 prl,                                                                                                                     
]{revtex4-1}                                                                                                              
                                                                                                                          
\usepackage[]{graphicx}
\usepackage{dcolumn}
\usepackage{bm}
\usepackage{hyperref}  
\hypersetup{%
colorlinks                                                                                                                
}   

\usepackage[hyperref,dvipsnames]{xcolor}
\usepackage{bbding}
 \usepackage{mathrsfs}
 \usepackage{amsbsy}
 \usepackage{empheq}
   
\usepackage{dashrule}
\usepackage{afterpage}

\newcommand\matlab[1]{}
\definecolor{Green}{RGB}{16,131,16} %

\newcommand\red[1]{\textcolor{black}{#1}}

\newcommand\ds[1]{\displaystyle{#1}}

\renewcommand\l{\ell}
 
\renewcommand\t{\ensuremath{\theta}}

\begin{document}

\title{Lubricated gravity currents of power-law
  fluids}

\author{P. Kumar}
 \affiliation{Ashoka University, Rajiv Gandhi Education City, Haryana 131029, India}
 \author{S. Zuri}
\affiliation{ Schulich Faculty of Chemistry, 
Technion, Haifa 3200003, Israel}
 \author{D. Kogan}
 \author{M. Gottlieb}
 \affiliation{Dept. of Chem. Eng., Ben-Gurion University of the
  Negev, Beer-Sheva 8410501, Israel}
\author{Roiy Sayag}\email{roiy@bgu.ac.il}                                                                                 
\homepage{http://www.bgu.ac.il/~roiy/}                                                                                    
  \affiliation{Dept. of Environ. Physics, BIDR, Ben-Gurion University of the Negev,
  Sde Boker 8499000, Israel}
\affiliation{Dept. of Mech. Eng., Ben-Gurion University of the
  Negev, Beer-Sheva 8410501, Israel}
\affiliation{Dept. of Physics, Ben-Gurion University of the
  Negev, Beer-Sheva 8410501, Israel}


%


%

\begin{abstract}
  The motion of glaciers over their bedrock or drops of fluid along a
  solid surface can vary dramatically when these substrates are
  lubricated.
  We investigate the coupled flow of a gravity current (GC) of
  strain-rate softening fluid that is lubricated by a denser,
  lower-viscosity Newtonian fluid.
  We present a set of experiments in which such GCs are discharged
  axisymmetrically and at constant flux over a flat surface.  Using
  imaging techniques we follow the front evolution of each fluid and
  their thickness field.
  We find that the two fronts of our lubricated GCs evolve faster than 
  non-lubricated GCs, though with similar time exponents. In addition,
  the thickness of the non-Newtonian fluid is nearly uniform while
  that of the lubricating fluid is nonmonotonic with localised
  spikes. Nevertheless, lubricated GCs remain axisymmetric as long as
  the flux of the lubricating fluid is sufficiently smaller than the
  non-Newtonian fluid flux.
\end{abstract}   

\maketitle

\begin{figure*}
    \hspace{-5mm}
    \includegraphics[width=16cm]{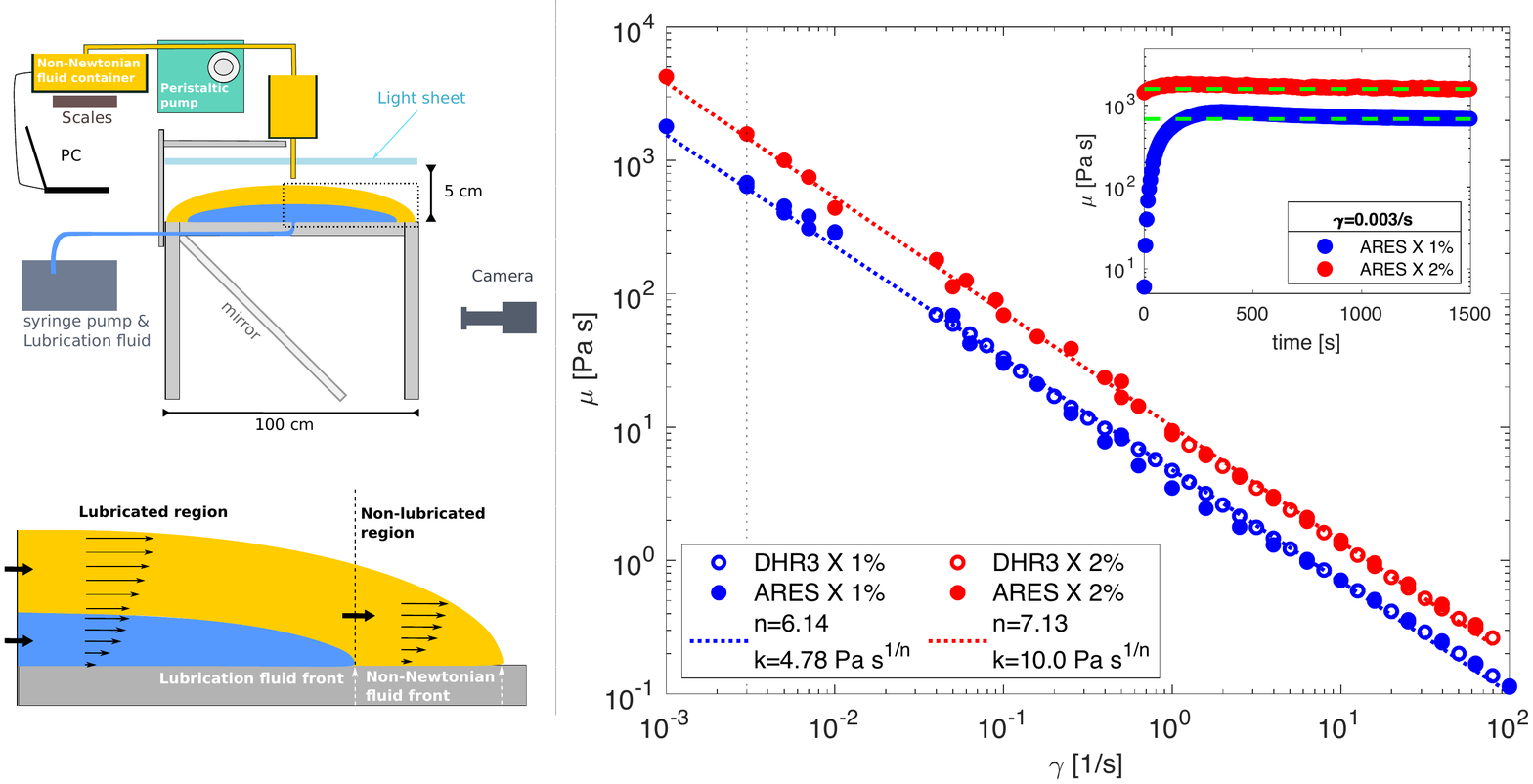}
  \begin{picture}(410,-20)(0,-10)
    \put(-20,185){$(a)$}
    \put(-20,105){$(b)$}
    \put(145,105){$(c)$}
    \put(45,155){$\rho,\mu(k,n)$}
    \put(35,140){$\rho_\l,\mu_\l$}
    \put(85,115){$r_L$}
    \put(120,115){$r_N$}
  \end{picture}
   \vspace{-38mm}  
   \caption{($a$) Our experimental apparatus for lubricated GCs.
     ($b$) Close-up of the flow region (dash-line rectangular in
     $(a)$). (c) Viscosity measurements of Xanthan solutions as a
     function of strain rate ($\bullet$, $\circ$) and the regression
     to power-law function
     (\tiny$\bullet$$\bullet$$\bullet$$\bullet$\small). Inset
     shows the evolution of the viscosity measurement at strain rate
     0.003/s. }
  \label{fig:schemes}
\end{figure*}

\section{Introduction}
Gravity-driven flows of one fluid over another can involve complex
interactions between the two fluids, which can lead to a rich dynamical
behavior.  Such flows occur in a wide range of natural and human-made
systems, as in lava flow over less viscous lava
\citep{Griffiths:2000-ARFM-dynamics,Balmforth:2000-Visco-plastic},
spreading of the lithosphere over the mid-mantle boundary
\citep{Lister-Kerr-1989:the,DauckBoxGellEtAl:2019--Shock}, 
ice flow over an ocean   
  \citep{DeContoPollard:2016--Contribution,KivelsonKhuranaRussellEtAl:2000--Galileo} and over bedrock
covered with sediments and water
\citep[][]{Stokes-Clark-Lian-et-al-2007:ice,Fowler:1987-theory}, flows
in porous media \citep{WoodsMason:2000-JFM-dynamics}, and droplets
motion on liquid-infused surfaces
\citep{KeiserKeiserClanetEtAl:2017-SM-Drop}.

The flow of GCs in circular geometry has been studied with a range of
boundary conditions. At the absence of a lubricating layer, a common
boundary condition along the base of the sole GC is no slip.  Such GCs
of Newtonian fluids that are discharged at constant flux follow a
similarity solution, in which the front position at time $t$ is
proportional to $t^{1/2}$ \citep{Huppert:1982-JFM-Propagation}.
Similar GCs of power-law (PL) fluids having exponent $n$, where $n=1$
represents a Newtonian fluid and $n>1$ represents a strain-rate
softening fluid, also have similarity solutions in which the front
propagation is proportional to $t^{(2n+2)/(5n+3)}$
\citep{Sayag-Worster:2013-Axisymmetric}.

On the other extreme, the presence of a lower fluid layer can
significantly reduce friction at the base of the top fluid, resulting
in extensionally dominated GCs. This is the case, for example, for ice
shelves, which deform over the relatively inviscid oceans with weak
friction along their interface.  The late-time front evolution of such
axisymmetric GCs of Newtonian fluids is proportional to $t$
\citep{PeglerWorster:2012-JFM-Dynamics}. However, when the top fluid
is strain-rate softening, an initially axisymmetric front can
destabilise and develop fingering patterns that consist of tongues
separated by rifts \citep{SayagWorster:2019-JFM-Instability1}.

In the more general case the conditions along the boundaries of the
GCs can vary spatiotemporally, as their stress field evolves.  For
example, the interface of an ice sheet with its underlying bed rock
can include distributed melt water and sediments, which impose
nonuniform and time-dependent friction along the ice base, and evolves
spatiotemporally under the stresses imposed by the ice layer.
Consequently, the coupled ice-lubricant system may evolve different
flow patterns
\citep[][]{Stokes-Clark-Lian-et-al-2007:ice,Fowler:1987-theory}.  Such
systems were modelled as two coupled GC of Newtonian fluids spreading
one on top of the other \citep{KowalWorster:2015-JFM-Lubricated}.  The
early stage of these flows follows a self similar evolution, in which
the fronts of the two fluids evolve like $t^{1/2}$, as in
non-lubricated (no-slip) GCs, and they can have radially non-monotonic
thickness.
However, experiments showed that at a later stage these coupled flows
became unstable and developed fingering patterns and non-axisymmetric
flow \citep{KowalWorster:2015-JFM-Lubricated}.

In this study we investigate a similar system as in
\cite{KowalWorster:2015-JFM-Lubricated}, only with a non-Newtonian
(strain-rate softening) GC as the top layer. We explore experimentally
the case where the flux and viscosity of the lubricating fluid is much
smaller than the non-Newtonian fluid. We trace the evolution of the
two front, and the spatiotemporal evolution of the light transmission
through the two fluids, from which we resolve the thickness
distribution of the top-layer, non-Newtonian fluid.  We then contrast
our findings with the present theories of non-lubricated and
lubricated GCs.

\section{Experimental setup}
\label{sec:experimental-setup}
 
The experimental apparatus (Figure \ref{fig:schemes}a,b) included a
flat, optically-transparent square glass sheet of 1x1 m$^2$ and 10 mm
thickness, supported by an aluminum frame parallel to the ground with
a 50 micron/m alignment accuracy. A rectangular plane mirror was
placed underneath the glass sheet at an angle 45$^\circ$ to the
horizontal for imaging. The center of the glass sheet had a 5 mm
diameter nozzle that was connected to a syringe pump (NE4000) that
delivered the lubricating fluid.
 The non-Newtonian fluid was driven by gravity from a beaker that was
 supported by an xyz-translation stage through an 8 mm diameter
 aluminium tube, whose outlet was 15 mm over the glass surface (Figure
 \ref{fig:schemes}a,b). We kept the flux constant by keeping a constant
 fluid level in the beaker using a peristaltic pump that supplied
 fluid from a reservoir.
 A 1x1 m$^2$ white light-sheet and a diffuser were positioned parallel
 to the glass surface and about 50 mm over it to illuminate the flow
 uniformly.  Time-lapsed image sequence was captured throughout each
 experiment using Nikon D5500 camera that was facing the 45$^\circ$
 mirror.

\begin{figure*}
  \vspace{0mm}
\begin{minipage}{0.4\textwidth}
  \vspace{0mm} 
  \begin{center}
    \renewcommand{\arraystretch}{1}
    \begin{tabular}{lccccccccc}
      \hline 
      \multicolumn{1}{c}{Experiment}
      & c &$Q$\\
                           \scriptsize{Non-lubricated GC}   &\%/w&g/s\\
      \hline
      GLSns1 ~\textcolor{yellow}{$\bullet$}                     &      1&1.673 \\ 
      GLSns2 ~\textcolor{yellow}{$\blacksquare$}                &      1&1.669 \\ 
      GLSs \textcolor{Orchid}{$\bigstar$}                         &    2&0.909 \\ 
      GLSns3 \textcolor{Orchid}{$\blacktriangleleft$}             &    2&0.917 \\ 
     ACRns1 \textcolor{Orchid}{\scriptsize$\mbox{\DavidStarSolid}$}                                                                  
                                                                   &    2&0.812 \\ 
      ACRns2 \textcolor{Orchid}{$\bullet$}                          &    2&0.811 \\ 
\hline
    \end{tabular} 
  \end{center}
  \includegraphics[width=4.9cm]{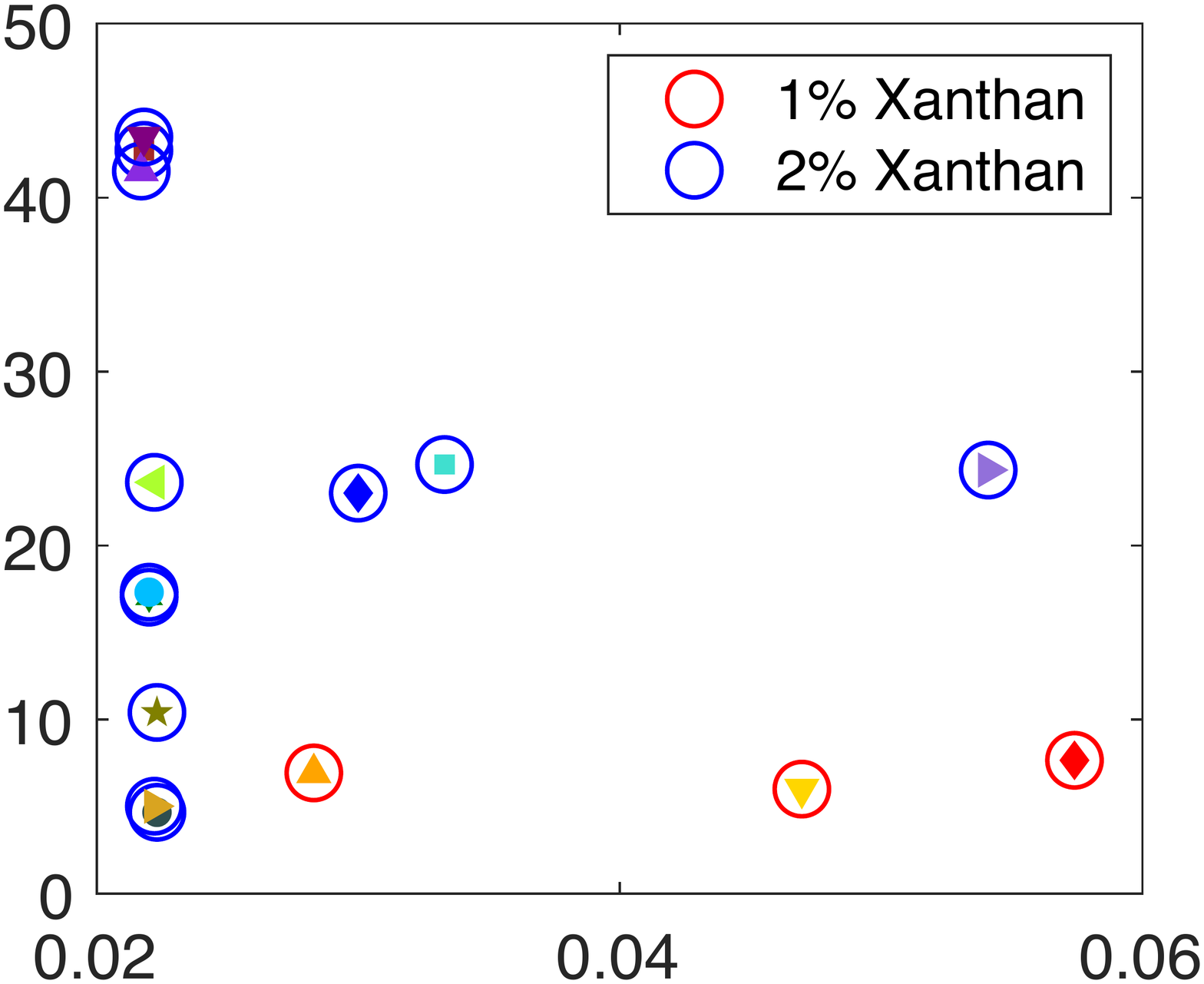}
  \begin{picture}(180,0)(-20,0)
    \put(70,3){$\mathscr{Q}$}
    \put(-5,60){$\frac{\mathscr{M}}{1000}$}  
  \end{picture}
\end{minipage}
\begin{minipage}{0.4\textwidth}
   \begin{center}
    \renewcommand{\arraystretch}{1}
    \begin{tabular}{lccccccccc}
      \hline 
      \multicolumn{1}{c}{Experiment} & {$\mathscr{Q}$} &{$\mathscr{M}$}
      & c &$Q$& $t_L$  & $r_N(t_L)$\\
                      \scriptsize{Lubricated GC}         & - & -  &\%/w&g/s& s & cm\\
      \hline
      1 ~\textcolor{red}{$\blacklozenge$}                    &0.0574 &    7661 &     1 &1.167   &  798   &  20  \\ 
      2 ~\textcolor{orange}{$\blacktriangle$}                &0.0283 &     6936 &     1 &1.767  & 752   &  20   \\ 
      3\footnote{Movie in supplementary material}\,\textcolor{yellow}{$\blacktriangledown$}            &0.04697&     6006 &      1&1.767  & 594   &  20   \\ 
      4$^{\textcolor{red}{a}}$\,\textcolor{PineGreen}{$\bullet$}  &0.0223 &    4671   &     2 & 4.716&   78  &  10   \\ 
      5 ~\textcolor{brown}{$\blacksquare$}                   &0.0218 &   42717   &     2 &0.643 &  1550  &  20  \\ 
      6 ~\textcolor{brown}{$\blacktriangleright$}            &0.0222 &    5023   &     2& 3.643 &   82   &  10  \\ 
      7$^{\textcolor{red}{a}}$\,\textcolor{SeaGreen}{$\blacktriangleleft$}          &0.0222 &   23639   &     2& 1.5   &  754   &  20  \\ 
      8 ~\textcolor{olive}{$\bigstar$}                       &0.0223 &   10412   &     2& 1.525 &  208   &  10  \\ 
     9 ~\textcolor{ForestGreen}{\scriptsize$\mbox{\DavidStarSolid}$}                                                                 
                                                            &0.022  &   17036    &   2 &1.5&     450    & 15   \\ 
      10 \textcolor{cyan}{$\bullet$}                         &0.022  &   17323    &   2 &1.5&     462    & 15   \\ 
      11 \textcolor{BlueGreen}{$\blacksquare$}               &0.0333 &   24661    &    2&1.5&     806   &  20   \\ 
      12 \textcolor{blue}{$\blacklozenge$}                   &0.03   & 23017      &  2  &1.664&  742     &20    \\ 
      13$^{\textcolor{red}{a}}$\textcolor{Fuchsia}{$\blacktriangle$}  &0.0217 &   41528    &    2&0.692&  1510   &  20   \\ 
      14 \textcolor{Plum}{$\blacktriangledown$}              &0.0218 &   43483    &    2&0.643&  1594   &  20   \\ 
      15 \textcolor{Orchid}{$\blacktriangleright$}           &0.0541 &   24343    &    2&1.533&    794   &  20  \\ 
\hline
    \end{tabular}
  \end{center}
\vspace{0mm}
\end{minipage}
%
\begin{picture}(0,0)(390,100)
  \put(0,0){(b)}
  \put(0,105){(a)}
  \put(155,0){(c)}
\end{picture}
\caption{
  (Table a) The non-lubricated GC experiments were performed on Glass
  (GLS) or Acrylic (ACR) substrates with a pre-coat of soap solution
  (s) or without one (ns). (b) The lubricated GC experiments in the
  $\mathscr{M}-\mathscr{Q}$ state space. (Table c) The lubricated GC
  experiments. All experiments were performed at 22$^\circ$C.}
  \label{ExTable}
\end{figure*}

\subsection{Preparation \& Properties of the experimental fluids}

The non-Newtonian fluid we used was an aqueous solution of food-grade
Xanthan gum (Jungbunzlauer) of 1\% and 2\% concentration (per weight).
To prepare a uniformly dissolved, and air-bubble-free solutions we
followed the following procedure: First, we generated a smooth vortex
(Eurostat-200) in a beaker containing deionized water. Then, we poured
the Xanthan-gum powder into the center of the vortex within less than
60s, to ensure uniform dispersion of the powder and prevent aggregates
before the viscosity of the solution soars. After 90 min of mixing we
added 0.3~g of lemon-yellow color per 5000 g of solution, and stirred
for additional 30 minutes. Finally, we stored the solution in a
4$^\circ$C refrigerator for 24h to remove air bubbles. The densities
of the solutions were 1.0035 and 1.007 gm/cm$^3$ for the 1\% and 2\%
concentrations, respectively.
  
The rheology of xanthan gum solutions of such concentrations has been
studied extensively in various flow configurations, and we describe it
in detail in \cite{SayagWorster:2019-JFM-Instability1}.  In general,
such xanthan solutions are viscoelastic. However, in flows such as the
one we consider here the  role of elastic deformation is
significantly smaller compared with viscous deformation
\citep[e.g.,][]{Sayag-Worster:2013-Axisymmetric,SayagWorster:2019-JFM-Instability1}. We
support this by estimating the Deborah number for our flow in \S
\ref{sec:discussion}.  Therefore, here we focus on the viscous
deformation of our xanthan solutions, which is known to be consistent
with a PL fluid of both shear and extensional thinning for a
wide range of strain rates, with an approximately similar exponent
\citep{MartinAlfonsoCuadriBertaEtAl:2018-CP-Relation}.

We used TA DHR-3 and ARES rheometers to measure the dependence of the
shear viscosity $\mu$ of our xanthan solutions on the shear rate
$\gamma$. The setup consisted of steady shear flow within a
2$^\circ$-cone-and-plate geometry (40~mm (DHR-3), and 50~mm (ARES))
and a solvent trap to minimize dehydration. The rate of shear was
varied monotonically in steps, and the measurement of the viscosity at
each step continued until a steady value was reached (within $\pm 1\%$
STD).  A weak-gel behaviour of the solution at low shear rates
resulted in increasingly longer viscosity measurements with the
decline in shear rate (Figure \ref{fig:schemes}c, inset). Fitting our
measured viscosity to the power-law relation
\begin{equation}
  \mu=k \gamma^{{1/ n}-1}, \nonumber
\end{equation}
with the consistency $k$ and the exponent $n$ as free parameters, we
find that $n=6.14,~7.14$ and $k=4.78,~10.05$ Pa s$^{1/n}$ for the 1\%
and 2\% solutions, respectively  (Figure \ref{fig:schemes}c).

For a lubricating fluid of larger density and lower viscosity, we used
glucose solutions of density
$\rho_\l=1.164$~g/cm$^3$ and dynamic \red{viscosity of
  $\mu_\l=9$ mPa s}. We dyed the fluid in blue to get high contrast of
the evolving fluid-fluid interface, and to measure the evolution of
the non-Newtonian fluid thickness (\S\ref{sec:exp analysis})

\subsection{Experimental procedure \& preliminary experiments}

Two preparation procedures had a major impact on the reproducibility
of our experiments.
The first was the accuracy of concentric alignment of the two outlet
nozzles of the discharged fluids, which we did to an accuracy of
$\pm 100$ $\mu$m using the xyz stage and an adapter that could fit
simultaneously to both nozzles.
The second procedure concerns the impact of the wetting conditions
over the glass surface on the shape of the lubricating-fluid front.
At the absence of the non-Newtonian fluid, the axisymmetry of that
front was sensitive to the wetting conditions of the glass surface.
We avoided this sensitivity by forming uniform wetting conditions over
the glass surface using a coat of soap solution (3.3\% liquid dish 
soap in deionized water) that was allowed  to dehydrate prior to the
initiation of our experiments.
To verify that the soap coating had no impact on the non-Newtonian GC,
we compared the front evolution of a non-lubricated GCs with and
without a soap coating, and found no measurable difference for Xanthan
concentrations of either 1\% or 2\% (Figure \ref{ExTable}a).

\begin{figure*}
   \hspace{-10mm}
  \includegraphics[width=15.5cm]{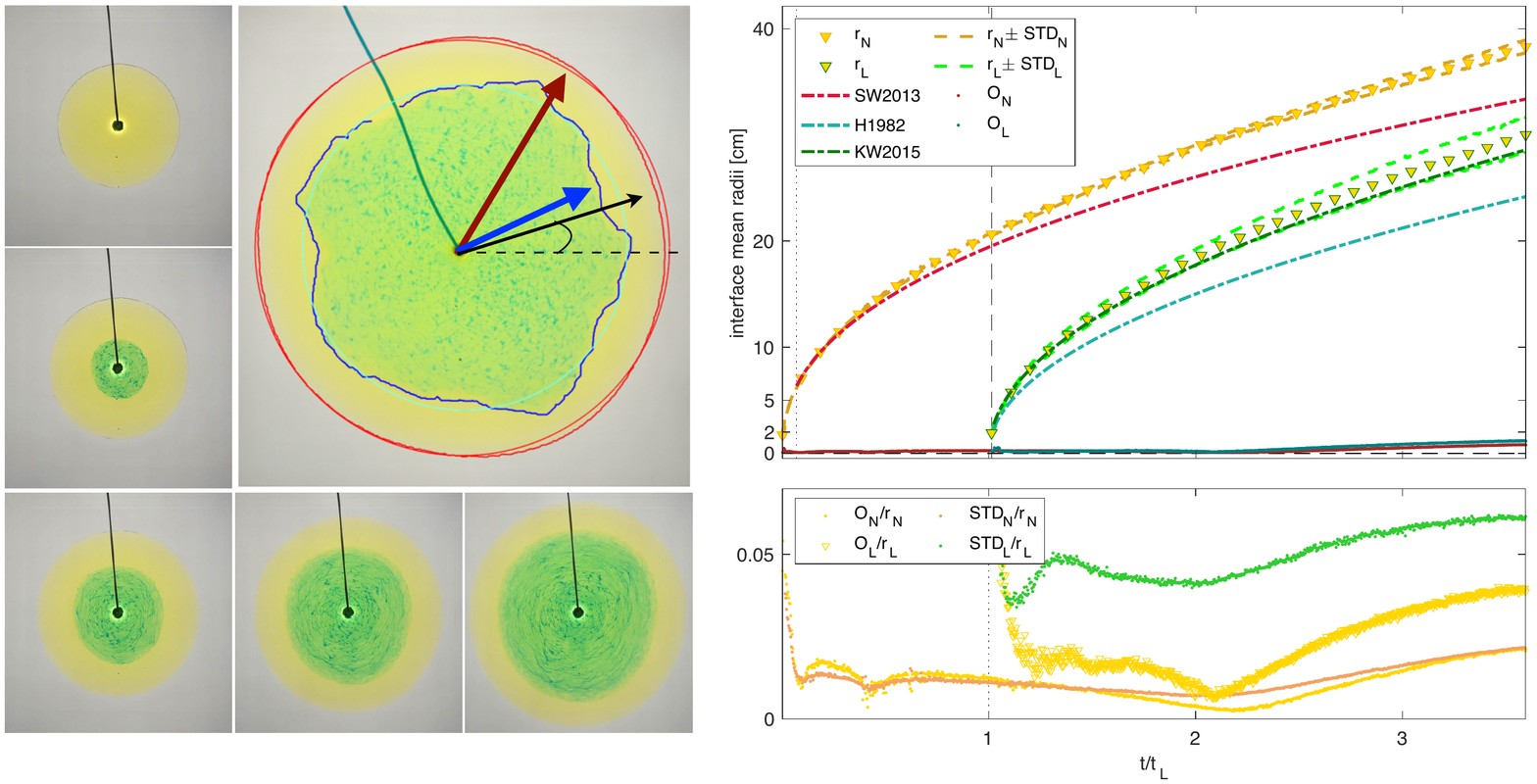}
  \begin{picture}(0,0)(418,-35)   
  \put(20,215){\color{black}$t=760$ s}
  \put(15,152){\color{black}$t=1010$ s}
  \put(15,91){\color{black}$t=1410$ s}
  \put(73,91){\color{black}$t=2010$ s}
  \put(133,91){\color{black}$t=2710$ s}
  \put(3,265){\textcolor{white}{$(a_i)$}}
  \put(3,203){\textcolor{white}{$(a_{ii})$}}
  \put(3,142){\textcolor{white}{$(a_{iii})$}}
  \put(61,142){\textcolor{white}{$(a_{iv})$}}
  \put(121,142){\textcolor{white}{$(a_{v})$}}
  \put(61,156){\textcolor{white}{$(b)$}}
  \put(182,150){$(c)$}
  \put(182,80){$(d)$}
  \put(122,245){$r_N$}
  \put(132,227){$r_L$}
  \put(147,213){$\theta$}
  \put(158,228){$r$}
\end{picture}
\vspace{-40mm}
\caption{(a) Time series of snapshots from experiment \#7,  showing the  Xanthan-solution  (yellow)
  lubricated by a sugar solution (green).
  $(b)$ Snapshot from experiment \#3 
  showing the resolved front of the lubricating
  (\textcolor{blue}{---}) and the non-Newtonian fluids
  (\textcolor{brown}{---}), and the fitted
  circles in (\textcolor{cyan}{---}) and (\textcolor{red}{---}),
  respectively.
  $(c)$  Evolution of the average front
  radii $r_N,r_L$,  centers $O_N,O_L$, and STDs compared with  theoretical  predictions of lubricated and
  non-lubricated GCs. 
  (d) STDs and centers $O_N,O_L$ normalised by instantaneous
  $r_N,r_L$. }
\label{fig:exp-seq} 
\end{figure*}

We conducted 15 lubricated GC experiments (Figure \ref{ExTable}c) in
the following procedure.  Initially, we released the non-Newtonian
fluid axisymmetrically in constant flux. When the evolving GC reached
a radius of 10-20 cm ($t=t_L$), we initiated the axisymmetric
discharge of the lubricating fluid underneath the non-Newtonian fluid
(Figure \ref{fig:exp-seq}a, Movies 1-4).

\section{Experimental analysis}
\label{sec:exp analysis}

We classify the lubricated GC experiments using the four dimensionless
numbers
\begin{eqnarray}
\ds{  \mathscr{Q}\equiv \frac{Q_{\l}}{Q},}&
  \ds{\mathscr{D}\equiv \frac{\rho_\l-\rho}{\rho},}&
    n,\nonumber\\
  \mathscr{M}\equiv\frac{\mu}{\mu_\l}=&\,\ds{\frac{\rho
    g}{\mu_\l}\left(\frac{k}{\rho
      g}\right)^{\frac{8n}{5n+3}}\left(Qt_L^{-4}\right)^{\frac{1-n}{5n+3}},}
\end{eqnarray}
representing, respectively, the sources flux ratio, the reduced
density ratio, the PL fluid exponent, and the viscosity ratio, in
which we evaluate the scale of $\mu$ using the characteristic scales
of a non-lubricated GC \citep{Sayag-Worster:2013-Axisymmetric}. In the
experiments we present, the lubrication flux was much smaller than the
flux of the PL fluid ($\mathscr{Q}\lesssim 0.06$), the viscosity ratio
varied over nearly one order of magnitude
($4700\lesssim\mathscr{M}\lesssim 43000$), and the 1\% and 2\% polymer
concentrations led to a small variation of $n$ (Figure
~\ref{fig:schemes}c) and to similar density ratios $\mathscr{D}_{2\%}=0.152$
and $\mathscr{D}_{1\%}=0.156$, respectively.

To explore the significance of the interaction between the two fluids,
we analyse the image sequence of each experiment to trace the
evolution of the two fluid fronts and fluid thicknesses, and we
compare the resulted measurements to existing theories of lubricated
and non-lubricated flows.

\subsection{The front evolution of the lubricating and non-Newtonian
  fluids.}
\label{sec:fronts}

Tracing the position of the fronts of the non-Newtonian fluid
$r_N(t,\t)$ and of the lubricating fluid $r_L(t,\t)$, where $\theta$
is the angular coordinate, we calculate the average radius of each
instantaneous front by fitting a circle (Figure~\ref{fig:exp-seq}b).
We find that the regression standard deviation (STD) can be reduced
significantly when using the centers of the fitted circles $O_N$ and
$O_L$ as additional free parameters (Figure
\ref{fig:exp-seq}c,d). This means that the deviation of the front from
an axisymmetric shape has two contributions: the centers $O_N, O_L$
that represent translation of the geometric center of the circular
front, and the STD that represents the non-axisymmetric displacement
of the front with respect to the translated circle.
We find that STD$_L, O_L\lesssim 5$~cm, whereas
STD$_N, O_N\lesssim 2$~cm, implying larger variations of the geometric
center and non-axisymmetric component of the lubricating fluid
compared with those of the non-Newtonian fluid. However, the ratio of
the centers and STDs with the front radii remain confined and even
decay (Figure \ref{fig:exp-seq}d). Therefore, throughout the flow, the
evolution of the fluid fronts remains axisymmetric to leading order.

\begin{figure*} 
   \hspace{-12mm}
   \includegraphics[width=18.cm]{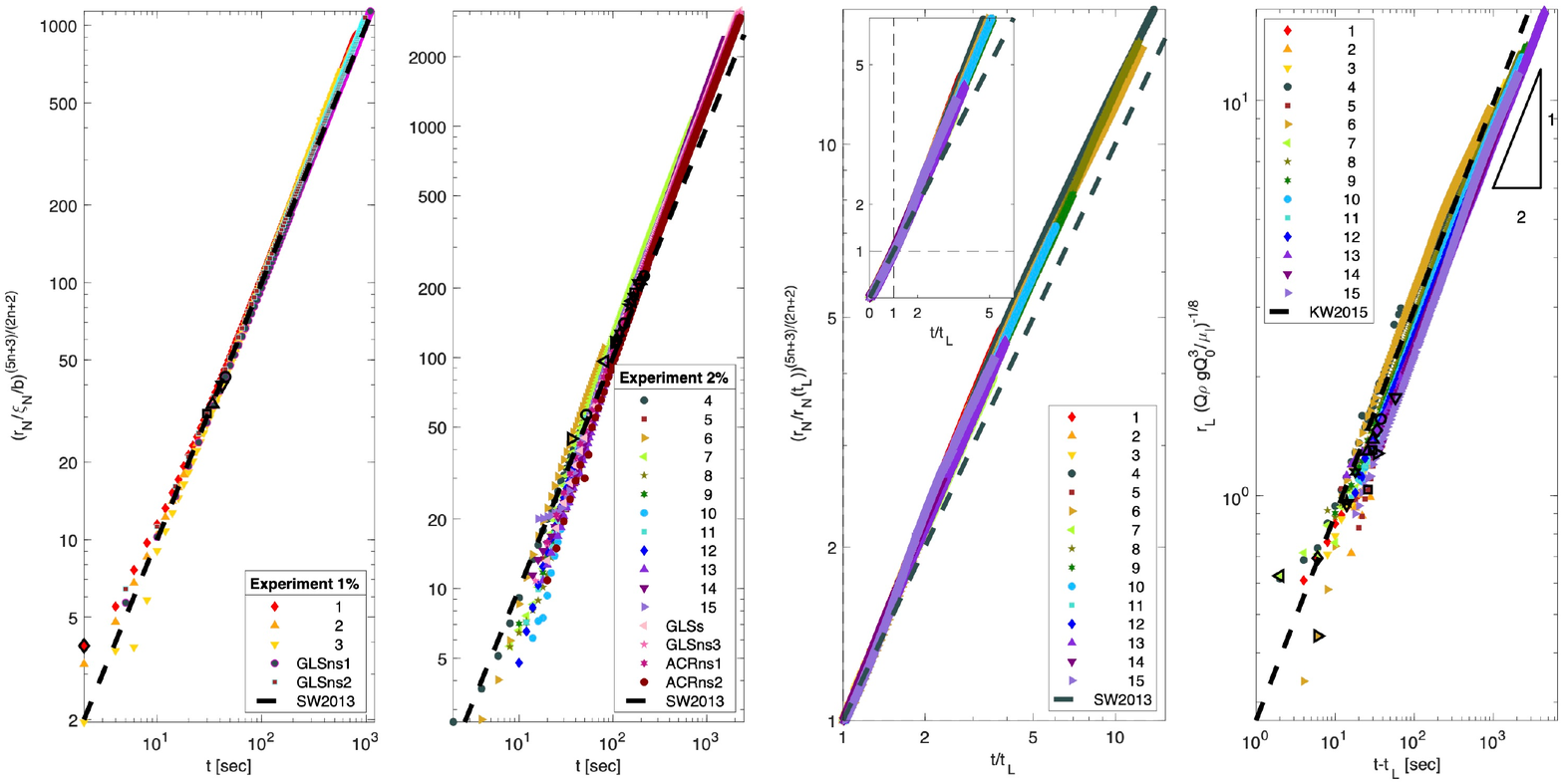} 
  \begin{picture}(0,0)(455,-135) 
    \put(-20,0){$(a)$}
    \put(90,0){$(b)$}
    \put(190,0){$(c)$}
    \put(304,0){$(d)$}
\end{picture}
\vspace{-46mm}
\caption{Evolution of the fluid fronts. (a, b) The front $r_N(t)$
  during the non-lubricated interval ($t<t_L$), for the 1\% (a) and
  2\% (b) concentrations.  (c) The front $r_N(t)$ during the
  lubricated interval ($t>t_L$). Inset zooms to the earlier stage that
  includes the non-lubricated interval. The corresponding front
  $r_N^{SW13}$ of the non-lubricated GC
  ($\textcolor{gray}{\hdashrule[0.5ex]{0.9cm}{1mm}{1.5mm}}$) has no
  fitting parameters (a,b,c).  (d) The front $r_L(t)$ compared with
  $r_L^{KW15}$
  ($\textcolor{black}{\hdashrule[0.5ex]{0.9cm}{1mm}{1.5mm}}$). 
  The markers when the thin-film criterion is first satisfied have
  black edges.}
\label{fig:rNearly-all} 
\end{figure*}

The resulted evolution of the fronts can be appreciated through
contrasting with existing theories of non-lubricated GC of PL
 \citep[][]{Sayag-Worster:2013-Axisymmetric}  and  Newtonian
 \citep[][]{Huppert:1982-JFM-Propagation} fluids, and
lubricated GC of Newtonian fluids
\citep[][]{KowalWorster:2015-JFM-Lubricated},
in which the
solutions for the fronts are respectively
\begin{subequations}
  \label{eq:similarity}
  \begin{align} 
    \label{SW13}
  r_{N}^{\textrm{SW13}}&=\xi_{N} b_N t^{{2n+2}\over{5n+3}},& b_N&=
  \left[{{2^{1-n}}\over{n+2}}Q^{2n+1}\left({\rho g}\over{k}\right)^{n}
                                                  \right]^{1/(5n+3)},\\
    \label{H82}
    r_{L}^{\textrm{H82}}&=\xi_{L} b_L t^{1/2},& b_L&=
                                      \left(Q_\l^{3}({\rho_\l-\rho})g\over{3\mu_\l} \right)^{1/8},
    \\    \label{KW15} 
    r_{L}^{\textrm{KW15}}&=\xi_L^*b_L^*t^{1/2},& b_L^*&=\left(\mathscr{MQ}Q^3{\rho g}\over{\mu}\right)^{1/8}=\left(\mathscr{Q}Q^3{\rho g}\over{\mu_\l}\right)^{1/8},
  \end{align}
\end{subequations}
where for constant flux $\xi_N=\xi_L\approx 0.71$  and
 $\xi_L^*\approx 0.27\pm 0.02$. These similarity solutions
 are valid when the thin-film approximation is satisfied, i.e., when
 $r_N/h\gtrsim 10$, where $h$ is the characteristic
 thickness. 
 We estimate this ratio by assuming that the instantaneous fluid
 volume $Qt$ is distributed as a disc of radius $r_N$ and thickness
 $h$. 
 Therefore, the condition is $r_N/h=\pi r_N^3/Qt\gtrsim 10$ for the
 top fluid, and similarly, $r_L/h_\l=\pi r_L^3/Q_\l(t-t_L)\gtrsim 10$
 for the lubricating fluid.

 We consider the evolution of the front $r_N$ of the non-Newtonian
 fluid separately during the non-lubricated interval ($t<t_L$, Figure
 \ref{fig:rNearly-all}a,b), and the lubricated interval (Figure
 \ref{fig:rNearly-all}c). In the first interval we expect the front to
 evolve consistently with \eqref{SW13} as soon as the
 lubrication-approximation condition is satisfied. We find that our
 measurements for the 1\% solutions indeed collapse to the theoretical
 predictions (Figure \ref{fig:rNearly-all}a).  For the 2\% solutions
 our measured front propagates faster than the theoretical prediction
 (Figure \ref{fig:rNearly-all}b).  Our preliminary experiments
 (Figures \ref{ExTable}a, \ref{fig:rNearly-all}a,b) indicate that this
 discrepancy is independent of the substrate wetting conditions or the
 substrate material (glass or acrylic). Therefore, we believe that wall
 slip may arise at high polymer concentrations, as we elaborate in
 \S\ref{sec:discussion}.
 In the lubricated interval ($t\ge t_L$), we measure faster
 propagation of $r_N$ than a non-lubricated GC (Figure
 \ref{fig:rNearly-all}c), which implies a significant impact of
 lubrication. The response of $r_N$ to the lubrication fluid
 initialises with a transient acceleration
 ($1.5\lesssim t/t_L\lesssim 5$) with respect to the non-lubricated
 solution, then ($5\lesssim t/t_L$) the front evolves with a similar
 exponent as the non-lubricated solution (eq. \ref{SW13}), but faster.

The flow of the lubricating fluid is affected by both the normal and
shear stresses applied by the surrounding non-Newtonian
fluid. Nevertheless, we find that the evolution of $r_L$ is power law
in time with an exponent 1/2 (Figure \ref{fig:rNearly-all}d), similar
to the exponent predicted for Newtonian Ks
\citep{Huppert:1982-JFM-Propagation,KowalWorster:2015-JFM-Lubricated}. However,
our measured intercept deviates from the theoretical prediction
(Figure \ref{fig:rNearly-all}d), implying that it may have dependence
on the properties of the non-Newtonian fluid, which is absent in the
intercept $b_L^*$ of the Newtonian theory \eqref{KW15}.
We also note that for all the experiments
\begin{equation}
  \label{eq:intercept_ratio}
  b_L/b_L^*=\left(\tfrac{1}{3}\mathscr{D}\mathscr{Q}^2\right)^{1/8}<1,
\end{equation}
implying that the interaction with the top fluid layer has a significant
impact on the propagation of the lower-layer front $r_L$, as can be
appreciated in Figure \ref{fig:exp-seq}c.

\subsection{The thickness evolution of the lubricated power-law fluid}

\begin{figure*} 
   \hspace{-12mm}
  \includegraphics[width=19.cm]{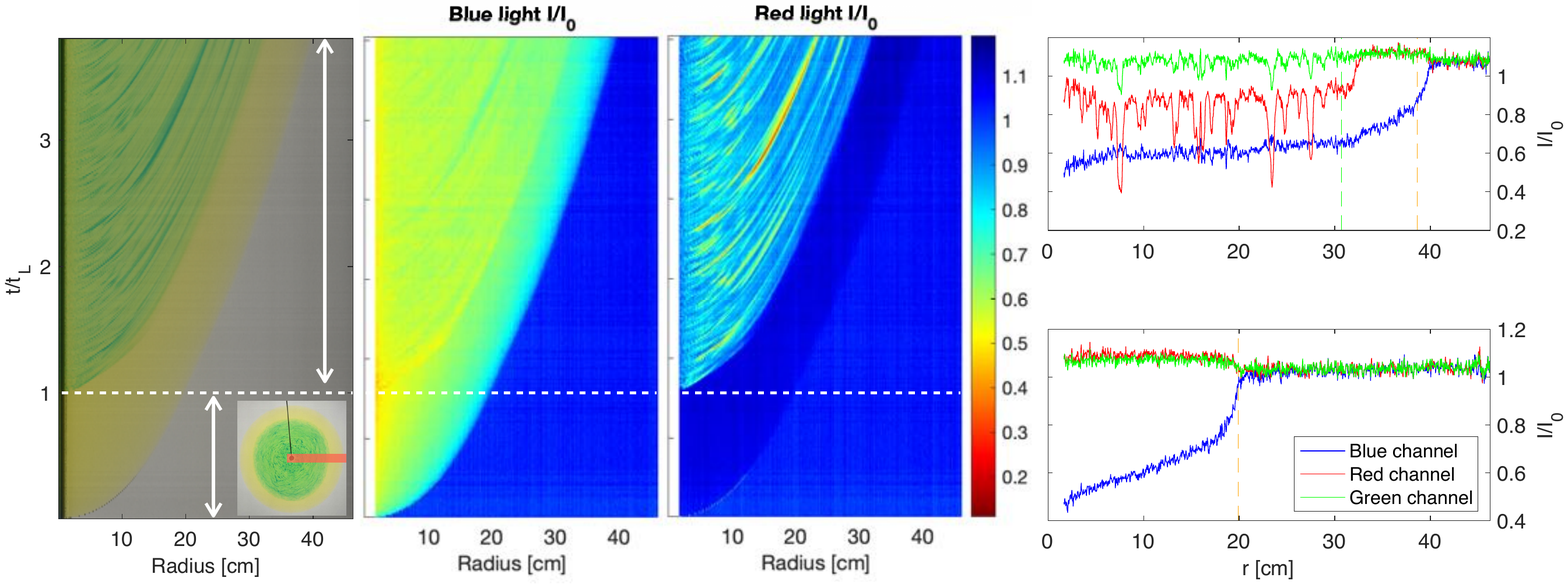}
  \begin{picture}(0,0)(200,-203)
    \put(-50,0){\textcolor{black}{$(a)$}}
    \put(55,0){$(b)$}
    \put(150,0){$(c)$}
    \put(265,0){$(d)$}
    \put(330,88){$t/t_L=1$} 
    \put(330,182){$t/t_L=3.8$}
    \put(39,75){\tiny\rotatebox{90}{\textcolor{white}{Lubricated}}}
    \put(-5,27){\tiny\rotatebox{90}{\textcolor{white}{Non-}}}
    \put(0,27){\tiny\rotatebox{90}{\textcolor{white}{lubricated}}}
\end{picture} 
\vspace{-72mm}
\caption{Tracing the fluid color intensity.  (a) Time series of
  snapshots along one radius (red line, inset).  (b) The normalised
  blue channel of the image in (a). (c) The normalised red channel of
  the image in (a). (d) The instantaneous red, green and blue
  intensities along one radius before lubrication ($t<t_L$, bottom),
  and after ($t>t_L$, top).}
\label{fig:intensity-all} 
\end{figure*}

\begin{figure*} 
   \hspace{-5mm}
  \includegraphics[width=18.cm]{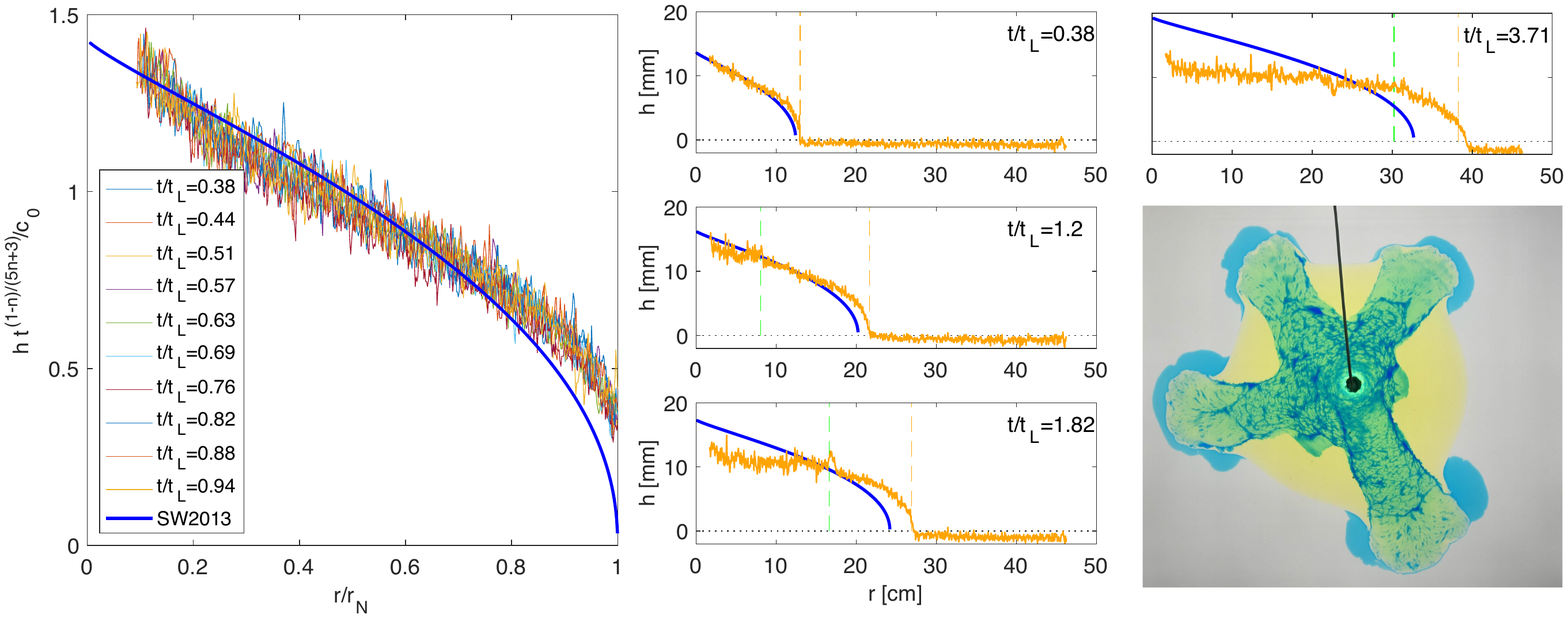}
  \begin{picture}(0,0)(450,-165)
    \put(-50,0){$(a)$}
    \put(145,0){$(b)$}
    \put(240,176){$(i)$}
    \put(378,176){$(iv)$}
    \put(235,116){$(ii)$}
    \put(235,57){$(iii)$}
    \put(295,0){$(c)$}
\end{picture}
\vspace{-53mm}
\caption{(a) regression of the blue-channel intensity during the
  non-lubricated interval ($t<t_L$) to the thickness solution of a
  non-lubricated GC (\textcolor{blue}{---}) to get
  $a_{c=\mbox{blue}}$. (b) Thickness evolution of a lubricated GC
  (experiment \#1, \textcolor{orange}{---}) and of the corresponding
  non-lubricated GC (\textcolor{blue}{---}).  Grid lines mark the
  measured $r_N$ (\textcolor{orange}{- - -}) and $r_L$
  (\textcolor{green}{- - -}). (c) High lubrication experiment
  ($\mathscr{Q}\approx 0.2,\mathscr{M}\approx 7500$) leads to
  emergence of radial streams.  }
\label{fig:thickness} 
\end{figure*}

Mass conservation implies that the thickness of lubricated GCs should
be smaller than that of non-lubricated GCs to account for the
difference with their faster front propagation. To investigate this we
measure the fluid thickness field from the projected images of the GC.
When a monochromatic light of intensity $I_0$ propagates through a
fluid layer, the intensity of the transmitted light $I$ drops
exponentially with the fluid thickness $h$ according to the
Beer-Lambert-Bouguer law \citep[e.g.,][]{Bouguer:1729-PFCJ-Essai}
\begin{equation}
  \label{eq:BL-law}
  h=-\frac{1}{a_c}\ln(I/I_0),
\end{equation}   
where $a_c$ is the attenuation coefficient, which we consider as depth
independent. We calculate $a_c$ using the normalised intensity $I/I_0$
of a non-lubricated GC together with the known solution for the
thickness of a non-lubricated GC of PL fluid
\begin{equation}
  \label{eq:hSW2013}
  h=c_0(Q,\rho,n)t^{(n-1)/(5n+3)}\psi(r/r_N),
\end{equation}
where $c_0$ is a known constant and $\psi$ is the dimensionless
solution to a nonlinear differential equation that we solve
numerically \citep{Sayag-Worster:2013-Axisymmetric}. Specifically, we
trace the transmitted light intensity along a radius during the
non-lubricated part of the flow, when only the PL fluid is present
(Figure \ref{fig:intensity-all}a) to get $I(r,t)/I_0$ (Figure
\ref{fig:intensity-all}b,c).
The yellow dye of the PL fluid absorbs the blue and hardly attenuates
the red and green wavelengths. Therefore, we expect a growing attenuation of the
blue component with the non-Newtonian fluid thickness (Figure
\ref{fig:intensity-all}d).
Consequently, fitting the blue light normalised transmission
$I(r,t<t_L)/I_0$ to \eqref{eq:BL-law}, where eq. \eqref{eq:hSW2013} is
substituted for the thickness (Figure \ref{fig:thickness}a), we obtain
the blue light attenuation coefficient
$a_{c=\mbox{blue}}\approx 0.05~$mm$^{-1}$.

Having $a_{c=\mbox{blue}}$ allows us to trace the thickness of the PL
fluid also where it is over a layer of lubrication fluid.  This can be
done even though the light is transmitted through two layers of
different fluids because the lubrication fluid is dyed in blue,
implying that the blue wavelength is transmitted through it with
negligible attenuation and that nearly all of the blue light
attenuation occurs in the yellow top layer. Consequently, we can infer
the thickness of the PL fluid throughout the flow using
\eqref{eq:BL-law} with the same coefficient $a_{c=\mbox{blue}}$
(Figure \ref{fig:thickness}b). We find that once the lubricating fluid
is discharged (Figure \ref{fig:thickness}b$_{(ii-iv)}$) the thickness
of the top layer drops over time with respect to that of the
non-Newtonian GC, as we expect, and becomes nearly uniform along the
entire lubricated part.

The thickness of the lubricating layer can, in principle, be
calculated using the same technique, but using the red component of
the transmitted light, which is absorbed mostly by the blue-dyed
lubricating fluid (Figure \ref{fig:intensity-all}c,d). However, at the
absence of a theory for lubricated GC of PL fluids, we cannot apply
with confidence the above technique to convert red light intensity to
lubricating-layer thickness. 
Alternatively, such calibration can be achieved by measuring light
transmission through a fluid layer of known thickness
\citep[][]{VernayRamosLigoure:2015-JoFM-Free}. Here we estimate the
thickness of the lubricating film to get a sense of scale.  Assuming
the lubricating fluid has a uniform thickness $h_\l$, conservation of
mass implies that $h_\l=Q_\l (t-t_L)/\rho_\l\pi r_L^2$. Using the
parameters of experiment \#1 (Figure \ref{ExTable}) and the measured
$r_L(t)$ we find that at $t/t_L=1.2$ ($r_L=8.05$~cm)
$h_\l\approx 0.437$~mm (Figure
\ref{fig:thickness}b$_{ii}$), 
and at $t/t_L=3.71$ ($r_L=30.21$~cm) $h_\l\approx 0.434$~mm (Figure
\ref{fig:thickness}b$_{iv}$).
This implies that the thickness of the lubricating-fluid layer in that
experiment is approximately 25 times thinner than that of the PL fluid
layer. Therefore, the thickness profiles of the non-Newtonian fluid in
Figure \ref{fig:thickness}b represent quite accurately the free
surface of the GC.

\section{Discussion}
\label{sec:discussion}

We assumed that in the flow experiments the dominant deformation
mechanism of our xanthan solutions is viscous. To confirm this we
estimate the Debora number $De=\lambda/T$, where
$\lambda\approx 0.1\,$s is the elastic relaxation time for 1\%
solutions
\citep[][]{StokesMacakovaChojnicka-PaszunEtAl2011--Lubrication} and
$T=\pi r^2 h\rho/Q$ is the characteristic timescale of the flow.
Having $h\approx 1$~cm, $Q/\rho\approx 0.5-5$~cm$^3$/s, and using
$r_N(t_L)\approx 10-20$~cm for the radius (Figure \ref{ExTable}), the
smallest timescale we obtain is $T\approx 60$~s, implying
$De\approx 0.0017$ and negligible elastic deformation.  The radius
where elasticity may become important ($De\approx 1$) is
$r\approx 4$~mm, which is the radius of the nozzle.  The relaxation
time of 2\% solutions is larger than $\lambda$, but even if it is
tenfold larger, $De$ would still be low.

During the non-lubricated interval we measure faster evolution of the
front of the 2\% Xanthan solution than the prediction of
\cite[][]{Sayag-Worster:2013-Axisymmetric}.  This inconsistency may
arise due to a more intense polymer entanglement in the 2\% solution
than in the 1\% solution \citep[the storage modulus of the former is
$\sim 3$ times larger than of the
latter,][]{MartinAlfonsoCuadriBertaEtAl:2018-CP-Relation}.
Consequently, one process that could lead to a faster propagation and
be more substantial in the 2\% solutions is wall slip at the
fluid-solid interface, potentially through adhesive failure of the
polymer chains at the solid surface or through cohesive failure due to
disantanglement of chains in the bulk from chains adsorbed at the wall
\citep{BrochardGennes:1992-L-Shear}. Such mechanisms would turn the
theoretically assumed no-slip condition invalid.

Although we did not calculate the thickness of the lubricating fluid,
the true-color images and the transmitted red-light intensity imply
that it has a complex spatiotemporal structure. In particular, the
thickness is non-monotonic, containing spikes of 50\% reduction in the
transmission that appear to move downstream with the flow (Figure
\ref{fig:intensity-all}). This structure may arise due to roughness of
the non-Newtonian fluid free surface, which may imply the presence of
a yield stress. Alternatively, an instability of the fluid-fluid
interface may lead to the growth of finite-amplitude thickness spikes.

The transition to the lubricated interval coincides with the growth of a
non-axisymmetric component of the fronts and translation of the GCs
geometric center. However, as the GC evolves these quantities appear
to become confined and attenuated (Figure \ref{fig:exp-seq}d),
implying that the flow remains axisymmetric to leading order. This
apparent stability may be associated with the substantially lower
lubrication flux compared with the flux of the top non-Newtonian
fluid. Our hypothesis is reinforced by preliminary experiments with
larger $\mathscr{Q}$, which indicate breakdown of axisymmetry and the
development of finger-like patterns in both the lubricating and
non-lubricating flows (Figure \ref{fig:thickness}c).

\section{Conclusions}

Lubricated GCs are controlled by complex interactions between two
fluid layers. The lower, lubrication layer modifies the friction
between the substrate and the top layer, which in turn applies
stresses that affect the distribution of the lubricating layer.  The
resulted flow can vary dramatically from non-lubricated GCs.

In our experiments at low flux ratio ($\mathscr{Q}<0.06$) we find that
the two fronts follow a power-law evolution with similar exponents as
their corresponding non-lubricated GCs, but they propagate faster than
the fronts of non-lubricated GC.
Specifically, following a transient acceleration during the lubricated
stage, the front $r_N$ of the non-Newtonian fluid evolves with the
same exponent $(2n+2)/(5n+3)$ as a non-lubricated PL fluid, but
with a larger intercept.
The front $r_L$ of the lower lubricating fluid evolves with the same
exponent $1/2$ as Newtonian GCs, but with an intercept that appears
sensitive to the properties of the overlaying non-Newtonian fluid.
Unlike non-lubricated GCs, the radial distribution of the thickness of
the non-Newtonian fluid in the lubricated region is nearly uniform,
whereas that of the lubricating fluid appears to be nonmonotonic with
localised spikes.  
Nevertheless, the fronts of the two fluids remain axisymmetric
throughout the flow, with a weak non-axisymmetric component that
relaxes or remains confined. 

We believe that axisymmetric models can provide accurate predictions
of these flows, to leading order. However, preliminary experiments
with $\mathscr{Q}>0.06$ reveal strong symmetry breaking of the
circular fronts and a transition to stream-like flow, implying that
for a sufficiently large $\mathscr{Q}$ the axisymmetric fronts can
turn unstable.  The experiments and analysis we present may provide
new insights into the rich phenomenology in potentially dynamically
similar systems such as ice-sheet systems coupled with hydrological
networks.

\begin{acknowledgments}       
We thank D. Bokobza, S. Kabalo and V. Melnichak for their assistance
in setting up the experimental apparatus, and to
\textit{Jungbunzlauer} for the Xanthan gum. This research was
supported by the GERMAN-ISRAELI FOUNDATION (grant No. I240430182015).
Declaration of Interests. The authors report no conflict of interest.
\end{acknowledgments}

\bibliographystyle{apsrev4-1}  
 
 %

\end{document}